\documentclass[traditabstract]{aa}

\usepackage{graphicx}
\usepackage{amssymb}

\begin{document}

\title{Orbital evolution under action of fast interstellar gas flow}
\author{P. P\'{a}stor$^{1,2}$ \and J. Kla\v{c}ka$^{1}$
\and L. K\'{o}mar$^{1}$}
\institute{Department of Astronomy, Physics of the Earth, and Meteorology, \\
   Faculty of Mathematics, Physics and Informatics, \\
   Comenius University,
   Mlynsk\'{a} dolina, 842~48 Bratislava, Slovak Republic \\
   e-mail: pavol.pastor@fmph.uniba.sk, klacka@fmph.uniba.sk,
   komar@fmph.uniba.sk
   \and
   Tekov Observatory, \\
   Sokolovsk\'{a} 21, 934~01, Levice, Slovak Republic}
\date{}

\abstract
{
Orbital evolution of an interplanetary dust particle under action of an
interstellar gas flow is investigated. Secular time derivatives of the
particle orbital elements, for arbitrary orbit orientation, are presented.
An important result concerns secular evolution of semi-major axis.
Secular semi-major axis of the particle on a bound orbit decreases under
the action of fast interstellar gas flow. Possible types of evolution
of other Keplerian orbital elements are discussed. The paper compares
influences of the Poynting-Robertson effect, the radial solar wind and
the interstellar gas flow on dynamics of the dust particle in outer
planetary region of the Solar System and beyond it, up to 100 AU.

Evolution of putative dust ring in the zone
of the Edgeworth-Kuiper belt is studied. Also non-radial solar wind
and gravitational effect of major planets may play an important role.
Low inclination orbits of micron-sized dust particles in
the belt are not stable due to fast increase of eccentricity caused
by the interstellar gas flow and subsequent planetary perturbations --
the increase of eccentricity leads to planet crossing orbits of the
particles.

Gravitational and non-gravitational effects are treated in a way which
fully respects physics. As a consequence, some of the published results
turned out to be incorrect. Moreover, the paper treats
the problem in a more general way than it has been presented up to now.

The influence of the fast interstellar neutral gas flow might not be
ignored in modeling of evolution of dust particles beyond planets.
}


\authorrunning{P. P\'{a}stor et al.}
\titlerunning{Orbital evolution under action of fast interstellar gas
flow}
\maketitle

\section{Introduction}

Motion of stars relative to their local interstellar medium
is frequent/usual process in galaxies. Neutral atoms penetrate
into the Solar System due to the relative motion of the Sun
with respect to the interstellar medium. This flow of neutral atoms
through a heliosphere has been investigated in many papers, e.g.
Fahr (1996), Lee et al. (2009), M\"{o}bius et al. (2009). Motion of dust
in interplanetary space can be affected by the neutral gas
penetrating into the heliosphere. Influence of this effect
on dynamics of dust particles is usually ignored in literature.
The Poynting-Robertson effect, the radial solar wind and the gravitational
perturbation of planet(s) are usually taken into account
(\v{S}idlichovsk\'{y} \& Nesvorn\'{y} 1994; Liou \& Zook 1997;
Liou \& Zook 1999; Kuchner \& Holman 2003).

Scherer (2000) has calculated secular time derivatives of angular
momentum and Laplace-Runge-Lenz vector of a dust particle under
the action of interstellar gas flow. But Scherer's calculations contain
several incorrectnesses. He has come to the conclusion that semi-major
axis of the dust particle increases exponentially (Scherer 2000, p. 334).
This paper presents that semi-major axis of the dust particle decreases
under the action of interstellar gas flow, in the framework of
the perturbation theory.

Motion of dust particles in the zone of the Edgeworth-Kuiper
belt under the action of the interstellar flow of gas has been investigated
by Kla\v{c}ka et al. (2009a). The authors have calculated secular time
derivatives of orbital elements only for the case when interstellar gas
velocity vector lies in the orbital plane of the dust
particle and direction of the velocity vector is parallel
with $y$-axis. This paper overcomes these restrictions.
Moreover, it presents some main properties of dust dynamics
under the action of the interstellar gas.

\section{Secular evolution}

Acceleration of a spherical dust particle caused by the flow of neutral
gas can be given in the form (Scherer 2000)
\begin{equation}\label{1}
\frac{d \vec{v}}{dt} = - ~c_{D} ~\gamma_{H} ~
\vert \vec{v} - \vec{v}_{H} \vert ~\left ( \vec{v} - \vec{v}_{H} \right ) ~,
\end{equation}
where $\vec{v}_{H}$ is velocity of the neutral hydrogen atom, $\vec{v}$
is velocity of the dust grain, $c_{D}$ is the drag coefficient,
$\gamma_{H}$ is the collision parameter.
For the collision parameter we can write
\begin{equation}\label{2}
\gamma_{H} = n_{H} ~\frac{m_{H}}{m} ~A ~,
\end{equation}
where $m_{H}$ is mass of the neutral hydrogen atom, $n_{H}$ is the
concentration of interstellar neutral hydrogen atoms, $A$ $=$ $\pi {R}^{2}$ is
the geometrical cross section of the spherical dust grain of radius $R$
and mass $m$. The concentration of interstellar hydrogen $n_{H}$ is not
constant in the entire heliosphere. For heliocentric distances $r$ less than
4 AU $n_{H}$ decreases precipitously from its value in the outer
heliosphere toward the Sun, due to ionization (Lee et al. 2009).
But in the outer heliosphere, $r$ $\in$ (30 AU, 80 AU), we can assume
that the concentration of the neutral hydrogen
atoms is constant $n_{H}$ $=$ 0.05 cm$^{-3}$ (Fahr 1996).
The same assumption can be used also behind
the solar wind termination shock. The shock was crossed by Voyager 1
at a heliocentric distance 94 AU and by Voyager 2 at 84 AU
(Richardson et al. 2008).

We will assume that the speed of interstellar gas is much greater than
the speed of the dust grain in the stationary frame
associated with the Sun ($v$ $\ll$ $v_H$). This approximation leads
to approximately constant value of $c_D$ $\approx$ 2.6
(Baines et al. 1965; Banaszkiewicz et al. 1994; Kla\v{c}ka et al. 2009a).

We want to find influence of the flow of interstellar gas on secular
evolution of particle's orbit. We will assume that the dust
particle is under the action of solar gravitation and the flow
of neutral gas. Hence we have equation of motion
\begin{equation}\label{3}
\frac{d \vec{v}}{dt} = - ~\frac{\mu}{r^{3}} \vec{r} ~-~ c_{D} ~\gamma_{H} ~
\vert \vec{v} - \vec{v}_{H} \vert ~\left ( \vec{v} - \vec{v}_{H} \right ) ~,
\end{equation}
where $\mu$ $=$ $G M_{\odot}$, $G$ is the gravitational constant,
$M_{\odot}$ is mass of the Sun, $\vec{r}$ is position vector
of the dust particle with respect to the Sun and
$r$ $=$ $\vert \vec{r} \vert$. The acceleration caused by
the flow of interstellar gas will be considered as a perturbation
acceleration to the central acceleration caused by the solar gravity.
In order to compute secular time derivatives of Keplerian orbital elements
($a$ - semi-major axis, $e$ - eccentricity, $\omega$ - argument of
perihelion, $\Omega$ - longitude of ascending node, $i$ - inclination)
we want to use Gauss perturbation equations of celestial mechanics.
To do this, we need to know radial, transversal and normal
components of acceleration given by Eq. (1). Orthogonal radial,
transversal and normal unit vectors of the dust particle
on the Keplerian orbit are (see Fig. 1 and e.g. P\'{a}stor 2009)
\begin{eqnarray}\label{4-6}
\vec{e}_{R} &=& \left (\cos \Omega ~\cos (f+ \omega) -
      \sin \Omega ~\sin (f+ \omega) ~\cos i ~, \right.
\nonumber \\
& &   \left. \sin \Omega ~\cos (f+ \omega) +
      \cos \Omega ~\sin (f+ \omega) ~\cos i ~, \right.
\nonumber \\
& &   \left. \sin (f+ \omega) ~\sin i \right ) ~,\\
\vec{e}_{T} &=& \left ( - \cos \Omega ~\sin (f+ \omega) -
      \sin \Omega ~\cos (f+ \omega) ~\cos i ~, \right.
\nonumber \\
& &   \left. - \sin \Omega ~\sin (f+ \omega) +
      \cos \Omega ~\cos (f+ \omega) ~\cos i ~, \right.
\nonumber \\
& &   \left. \cos (f+ \omega) ~\sin i \right ) ~,\\
\vec{e}_{N} &=& (\sin \Omega ~\sin i, ~- \cos \Omega ~\sin i, ~\cos i) ~,
\end{eqnarray}
where $f$ is true anomaly. Thus, we need to calculate the values of
$a_{R}$ $=$ $d \vec{v}/dt \cdot \vec{e}_{R}$,
$a_{T}$ $=$ $d \vec{v}/dt \cdot \vec{e}_{T}$ and
$a_{N}$ $=$ $d \vec{v}/dt \cdot \vec{e}_{N}$.
\begin{figure}[h]
\begin{center}
\includegraphics[height=0.30\textheight]{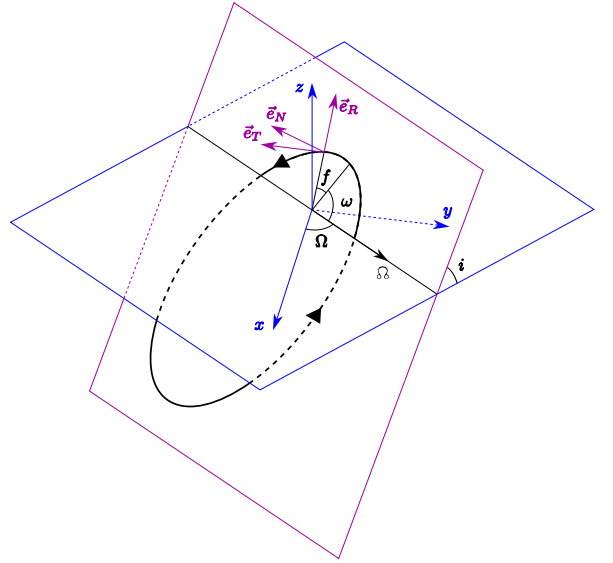}
\end{center}
\caption{A particle on an elliptical orbit depicted together with the radial,
transversal and normal unit vectors. Angles characterizing position of the
particle on the orbit are also shown.}
\label{F1}
\end{figure}
Velocity of the particle in an elliptical orbit can be calculated from
\begin{eqnarray}\label{7}
\vec{v} &=& \frac{d \vec{r}}{dt} = \frac{d}{dt} ~(r \vec{e}_{R})
\nonumber \\
&=&   r ~\frac{e \sin f}{1 + e \cos f} ~\frac{df}{dt} ~\vec{e}_{R} +
      r ~\vec{e}_{T} ~\frac{df}{dt} ~,
\end{eqnarray}
where
\begin{equation}\label{8}
r = \frac{p}{1 + e \cos{f}}
\end{equation}
and $p$ $=$ $a (1 - e^{2})$. In this calculation also the second
Kepler's law $df/dt$ $=$ $\sqrt{\mu p}/r^{2}$ must be used. Now, one can
easily verify that
\begin{eqnarray}\label{9-11}
(\vec{v} - \vec{v}_{H}) \cdot \vec{e}_{R} &=& v_{H} ~\sigma ~e ~\sin f -
      \vec{v}_{H} \cdot \vec{e}_{R}
\nonumber \\
&=&   v_{H} ~\sigma ~e ~\sin f - A ~,\\
      (\vec{v} - \vec{v}_{H}) \cdot \vec{e}_{T} &=& v_{H} ~\sigma ~
      (1 + e \cos f) - \vec{v}_{H} \cdot \vec{e}_{T}
\nonumber \\
&=&   v_{H} ~\sigma ~(1 + e \cos f) - B ~,\\
\nonumber \\
(\vec{v} - \vec{v}_{H}) \cdot \vec{e}_{N} &=& - ~
\vec{v}_{H} \cdot \vec{e}_{N} = - ~C ~,
\end{eqnarray}
where
\begin{equation}\label{12}
\sigma = \frac{\sqrt{\mu/p}}{v_{H}} ~,
\end{equation}
and $v_H$ $=$ $\vert \vec{v}_{H} \vert$.
Hence
\begin{eqnarray}\label{13}
\vert \vec{v} - \vec{v}_{H} \vert ^{2} &=& \frac{\mu}{p}(1 + 2 e \cos f +e^{2})
\nonumber \\
& &   - ~2 \sqrt{\frac{\mu}{p}} [B + e (A \sin f +B \cos f)] + v_{H}^{2} ~,
\end{eqnarray}
where the identity $\sqrt{A^{2}+B^{2}+C^{2}}$ $=$ $v_{H}$
was used.

If we denote components of the velocity vector of hydrogen gas in the
stationary Cartesian frame associated with the Sun as $\vec{v}_{H}$ $=$
$(v_{HX},v_{HY},v_{HZ})$, then we obtain
\begin{eqnarray}\label{14}
A \sin f + B \cos f &=& (-\cos \Omega ~\sin \omega
\nonumber \\
& &   - ~\sin \Omega ~\cos \omega ~\cos i) ~v_{HX}
\nonumber \\
& &   + ~(-\sin \Omega ~\sin \omega
\nonumber \\
& &   + ~\cos \Omega ~\cos \omega ~\cos i) ~v_{HY}
\nonumber \\
& &   + ~\cos \omega ~\sin i ~v_{HZ} = I ~.
\end{eqnarray}
Now we consider only such orbits for which
\begin{equation}\label{15}
\sigma \ll 1 ~,
\end{equation}
or, more precisely, the value $\sigma^{2}$ is negligible in comparison
with $\sigma$. This is reasonable for orbits with not very large
eccentricities, since $v$ $\ll$ $v_H$.
Using this approximation, Eqs. (13)-(14) yield
\begin{equation}\label{16}
\vert \vec{v} - \vec{v}_{H} \vert = v_{H}
[1 - \frac{\sigma}{v_H}(B + eI)] ~.
\end{equation}
For radial, transversal and normal components of acceleration we obtain from
Eq. (1), Eqs. (9)-(11) and Eq. (16)
\begin{eqnarray}\label{17-19}
a_{R} &=& - ~c_{D} ~\gamma_{H} ~v_{H}^{2} \Biggl [\frac{A}{v_{H}}
      \left (\frac{\sigma e I}{v_{H}} - 1 \right )
\nonumber \\
& &   + ~\sigma
      \left (e \sin f + \frac{AB}{v_{H}^{2}} \right ) \Biggr ] ~,\\
a_{T} &=& - ~c_{D} ~\gamma_{H} ~v_{H}^{2} \Biggl [\frac{B}{v_{H}}
      \left (\frac{\sigma e I}{v_{H}} - 1 \right )
\nonumber \\
& &   + ~\sigma
      \left (1 + e \cos f + \frac{B^{2}}{v_{H}^{2}} \right ) \Biggr ] ~,\\
a_{N} &=& - ~c_{D} ~\gamma_{H} ~v_{H} ~C
      \Biggl (\frac{\sigma e I}{v_{H}} - 1 +
      \sigma \frac{B}{v_{H}} \Biggr ) ~.
\end{eqnarray}
Now we can use Gauss perturbation equations of celestial mechanics
to compute time derivatives of orbital elements. The perturbation equations
have the form
\begin{eqnarray}\label{20}
\frac{d a}{d t} &=& \frac{2~a}{1~-~e^{2}} ~
      \sqrt{\frac{p}{\mu}} ~
      \left [ a_{R} ~e~ \sin f +
      a_{T} \left ( 1~+~e~ \cos f \right ) \right ] ~,
\nonumber \\
\frac{d e}{d t} &=&
      \sqrt{\frac{p}{\mu}} ~ \left [ a_{R} ~ \sin f + a_{T} \left ( \cos f ~+~
      \frac{e +	\cos f}{1 + e \cos f} \right ) \right ] ~,
\nonumber \\
\frac{d \omega}{d t} &=& -~ \frac{1}{e} ~ \sqrt{\frac{p}{\mu}} ~
      \left ( a_{R} \cos f - a_{T} ~
      \frac{2 + e \cos f}{1 + e \cos f} ~
      \sin f \right )
\nonumber \\
& &   - ~\frac{r}{\sqrt{\mu ~p}} ~
      a_{N} ~ \frac{\sin (f + \omega)}{\sin i} ~\cos i ~,
\nonumber \\
\frac{d \Omega}{d t} &=&
      \frac{r}{\sqrt{\mu ~p}} ~
      a_{N} ~ \frac{\sin (f + \omega)}{\sin i} ~,
\nonumber \\
\frac{d i}{d t} &=& \frac{r}{\sqrt{\mu ~p}} ~a_{N} ~ \cos (f + \omega) ~.
\end{eqnarray}
Time average of any quantity $g$ during one orbital period $T$ can
be computed using
\begin{eqnarray}\label{21}
\left \langle g \right \rangle &=& \frac{1}{T} \int_{0}^{T} g ~dt=
      \frac{\sqrt{\mu}}{2 \pi a^{3/2}} \int_{0}^{2 \pi} g
      \left (\frac{df}{dt} \right )^{-1} df
\nonumber \\
&=&   \frac{\sqrt{\mu}}{2 \pi a^{3/2}} \int_{0}^{2 \pi} g
      \left (\frac{\sqrt{\mu p}}{r^{2}} \right )^{-1} df
\nonumber \\
&=&   \frac{1}{2 \pi a^{2} \sqrt{1 - e^{2}}} \int_{0}^{2 \pi} g ~r^{2} ~df ~,
\end{eqnarray}
where the second ($\sqrt{\mu p}$ $=$ $r^{2} df/dt$) and the third
($4 \pi^{2} a^{3}$ $=$ $\mu T^{2}$) Kepler's laws were used.
From Eqs. (17)-(21) we finally obtain for the secular
time derivatives of the Keplerian orbital elements
\begin{eqnarray}\label{22-26}
\left \langle \frac{da}{dt} \right \rangle &=& - ~2 ~a ~c_{D} ~\gamma_{H} ~
      v_{H}^{2} ~\sqrt{\frac{p}{\mu}} ~\sigma ~
      \Biggl \{1 + \frac{1}{v_{H}^{2}}
\nonumber \\
& &   \times ~\Biggl [I^{2} - (I^{2} - S^{2})
      \frac{1 - \sqrt{1 - e^{2}}}{e^{2}} \Biggr ] \Biggr \} ~,\\
\left \langle \frac{de}{dt} \right \rangle &=& ~c_{D} ~\gamma_{H} ~
      v_{H} ~\sqrt{\frac{p}{\mu}} ~
      \Biggl [\frac{3I}{2} +
      \frac{\sigma (I^{2} - S^{2})(1 - e^{2})}{v_{H}e^{3}}
\nonumber \\
& &   \times ~\left (1 - \frac{e^{2}}{2} -
      \sqrt{1 - e^{2}} \right ) \Biggr ] ~,\\
\left \langle \frac{d \omega}{dt} \right \rangle &=& ~
      \frac{c_{D} ~\gamma_{H} ~v_{H}}{2} ~\sqrt{\frac{p}{\mu}} ~
      \Biggl \{ - ~\frac{3S}{e}
\nonumber \\
& &   + ~\frac{\sigma SI}{v_{H}e^{4}}
      \biggl [e^{4} - 6e^{2} + 4 -
      4(1 - e^{2})^{3/2} \biggr ]
\nonumber \\
& &   + ~C ~\frac{\cos i}{\sin i} ~
      \biggl [\frac{3e \sin \omega}{1 - e^{2}}
\nonumber \\
& &   - ~\frac{\sigma}{v_{H}}
      (S \cos \omega - I \sin \omega) \biggr ] \Biggr \} ~,\\
\left \langle \frac{d \Omega}{dt} \right \rangle &=&
      \frac{c_{D} ~\gamma_{H} ~v_{H} ~C}{2 \sin i} ~
      \sqrt{\frac{p}{\mu}} ~
      \Biggl [- ~\frac{3e \sin \omega}{1 - e^{2}}
\nonumber \\
& &   + ~\frac{\sigma}{v_{H}}(S \cos \omega - I \sin \omega) \Biggr ] ~,\\
\left \langle \frac{di}{dt} \right \rangle &=&
      - ~\frac{c_{D} ~\gamma_{H} ~v_{H} ~C}{2} ~
      \sqrt{\frac{p}{\mu}} ~
      \Biggl [\frac{3e \cos \omega}{1 - e^{2}}
\nonumber \\
& &   + ~\frac{\sigma}{v_{H}}(S \sin \omega + I \cos \omega) \Biggr ] ~,
\end{eqnarray}
where the quantities
\begin{eqnarray}\label{27}
S &=& (\cos \Omega ~\cos \omega -
      \sin \Omega ~\sin \omega ~\cos i) ~v_{HX}
\nonumber \\
& &   + ~(\sin \Omega ~\cos \omega +
      \cos \Omega ~\sin \omega ~\cos i) ~v_{HY}
\nonumber \\
& &   + ~\sin \omega ~\sin i ~v_{HZ} ~,
\nonumber \\
I &=& (- \cos \Omega ~\sin \omega -
      \sin \Omega ~\cos \omega ~\cos i) ~v_{HX}
\nonumber \\
& &   + ~(- \sin \Omega ~\sin \omega +
      \cos \Omega ~\cos \omega ~\cos i) ~v_{HY}
\nonumber \\
& &   + ~\cos \omega ~\sin i ~v_{HZ} ~,
\nonumber \\
C &=& \sin \Omega ~\sin i ~v_{HX} - \cos \Omega ~\sin i ~v_{HY} +
      \cos i ~v_{HZ} ~,
\end{eqnarray}
are values of
$A$ $=$ $\vec{v}_{H} \cdot \vec{e}_{R}$,
$B$ $=$ $\vec{v}_{H} \cdot \vec{e}_{T}$ and
$C$ $=$ $\vec{v}_{H} \cdot \vec{e}_{N}$ at perihelion of particle's
orbit ($f$ $=$ 0), respectively. The value of $C$ is a constant on
a given oscular orbit. The values of $S$, $I$ and $C$ are depicted in Fig. 2.
\begin{figure}[h]
\begin{center}
\includegraphics[height=0.15\textheight]{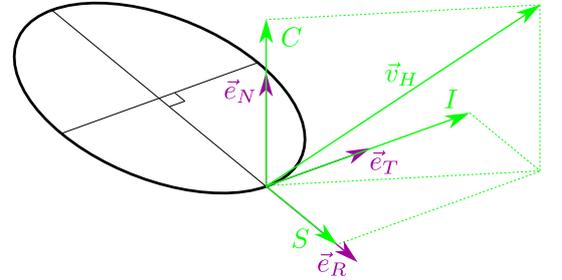}
\end{center}
\caption{A schematic representation of the values $S$, $I$ and $C$ for a given
orbit.}
\label{F2}
\end{figure}
One can use also the relations
\begin{eqnarray}\label{28}
S \cos \omega - I \sin \omega &=& \cos \Omega ~v_{HX} + \sin \Omega ~v_{HY} =
      \vec{e}_{PA} \cdot \vec{v}_{H} ~,
\nonumber \\
S \sin \omega + I \cos \omega &=& - \sin \Omega ~\cos i ~v_{HX}
\nonumber \\
& &   + ~\cos \Omega ~\cos i ~v_{HY}
\nonumber \\
& &   + ~\sin i ~v_{HZ} = \vec{e}_{PE} \cdot \vec{v}_{H} ~,
\nonumber \\
\vec{e}_{PA} \cdot \vec{e}_{PE} &=& 0 ~,
\nonumber \\
\vec{e}_{N} \times \vec{e}_{PA} &=& \vec{e}_{PE} ~.
\end{eqnarray}
Vector $\vec{e}_{PA}$ is directed from the Sun to
the ascending node. $\vec{e}_{PA}$ $\cdot$ $\vec{v}_{H}$ $=$
$S \cos \omega$ $-$ $I \sin \omega$ is magnitude of $\vec{v}_{H}$
component parallel with the line of nodes. The orbital plane is defined
by its normal unit vector $\vec{e}_{N}$. $\vec{e}_{N}$ $\times$
$\vec{e}_{PA}$ $=$ $\vec{e}_{PE}$. $\vec{e}_{PE}$
$\cdot$ $\vec{v}_{H}$ $=$ $S \sin \omega$ $+$ $I \cos \omega$ is magnitude
of $\vec{v}_{H}$ component perpendicular to the line of nodes
and lying in the orbital plane.

\begin{figure*}[t]
\begin{center}
\includegraphics[width=0.95\textwidth]{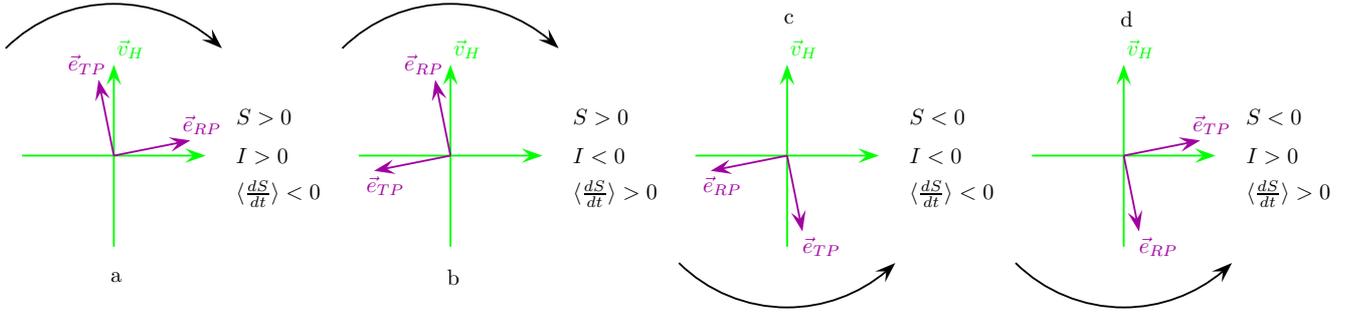}
\end{center}
\caption{Secular time derivatives of $S$ (see Eqs. 27) for a dust
particle on prograde orbit in planar case. Origins of these
Cartesian coordinate systems are in the Sun and vertical axes are aligned with
the direction of the hydrogen gas velocity vector. $\vec{e}_{RP}$ and
$\vec{e}_{TP}$ are unit radial and transversal vectors in perihelion of
the particle orbit (see text).}
\label{F3}
\end{figure*}

\section{Theoretical discussion}

Eqs. (22)-(26) enable to deduce some properties of secular
evolution of the dust particle under the action of the flow
of interstellar gas. $C$ $=$ 0 for a special case when
the velocity of hydrogen gas $\vec{v}_{H}$ lies in the orbital
plane of the particle. In this case we get that the inclination
and the longitude of ascending node are constant. Secular time derivatives
for a special case of this kind ($i$ $=$ $\Omega$ $=$ 0,
$v_{HX}$ $=$ $v_{HZ}$ $=$ 0, $v_{HY}$ $=$ $v_{H}$ in Eq. 27)
have been derived in Kla\v{c}ka et al. (2009a).
Secular time derivatives of $a$, $e$ and $\omega$ are
given in Kla\v{c}ka et al. (2009a) without generalization
represented by Eqs. (27) and Fig. 2.

Putting $\sigma$ $=$ 0 in Eqs. (22)-(26)
one obtains a solution equivalent to the solution
of Eq. (1) with the RHS independent of the particle's
velocity (constant force). Eq. (22) yields
constant semi-major axis for the unrealistic case $\sigma$ $=$ 0.
We will show, using Eq. (22), that secular semi-major axis
decreases under the action of interstellar gas flow
for $\sigma$ $>$ 0. If secular increase
of the semi-major axis would occur, then the value of curl
braces in Eq. (22) would be negative. The value of $1 - \sqrt{1 - e^{2}}$
is always positive or zero. Thus, the curly braces could be
negative only for $I^{2} - S^{2}$ $>$ 0.
Since $I$ and $S$ are radial and transversal components of the constant
vector $\vec{v}_{H}$ in the perihelion (Fig. 2) of particle's orbit, we
obtain maximal value of $I^{2} - S^{2}$ for orbit orientation
characterized by $S$ $=$ 0. Using these assumptions, the minimal value
($MV$) in the curl braces is
\begin{eqnarray}\label{29}
MV &=& 1 + \frac{1}{v_{H}^{2}}
      \left (I^{2} - I^{2}\frac{1 - \sqrt{1 - e^{2}}}{e^{2}} \right )
\nonumber \\
&=&   1 + \frac{I^{2}}{v_{H}^{2}}\frac{e^{2} - 1 + \sqrt{1 - e^{2}}}{e^{2}}
\nonumber \\
&=&   1 + \frac{I^{2}}{v_{H}^{2}} \sqrt{1 - e^{2}} ~
      \frac{1 - \sqrt{1 - e^{2}}}{e^{2}} > 0 ~.
\end{eqnarray}
The positiveness of $MV$ means that the secular semi-major axis is
a decreasing function of time. Eq. (22) was derived under the assumption
that Solar gravity represents dominant acceleration in comparison to
the interstellar gas disturbing acceleration. Since gravitational acceleration
decreases with square of a heliocentric distance, the assumption about
the disturbing interstellar gas acceleration may not be valid far from the Sun.
Behind the heliocentric distance at which solar gravitational acceleration
equals the interstellar gas acceleration, the particle's secular semi-major
axis can also be an increasing function of time. The two accelerations
are equal at heliocentric distance $\approx$ 1 $\times$ 10$^{4}$ AU
for dust particle with $R$ $=$ 1 $\mu$m and $\varrho$ $=$ 2 g/cm$^{3}$,
and, at heliocentric distance $\approx$ 7 $\times$ 10$^{3}$ AU
for dust particle with $R$ $=$ 1 $\mu$m and $\varrho$ $=$ 1 g/cm$^{3}$.
If Eq. (22) can be used, then secular time derivative of the semi-major
axis is proportional to the value of semi-major axis (value of
$\sqrt{p/ \mu} ~\sigma$ is independent of semi-major axis).
Our result differs from the Scherer's statement (semi-major of
the particle can increase exponentially) at least due to an error
in his expression for secular time derivative of magnitude
of angular momentum.

Defining the function $w (e)$ $\equiv$ 1 $-$ $e^{2}$ / 2 $-$
$\sqrt{1 - e^{2}}$, $e$ $\in$ [0,1), present in Eq. (23), we obtain
\begin{eqnarray}\label{30}
\frac{dw}{de} &\equiv& \frac{d}{de}
      \left (1 - \frac{e^{2}}{2} - \sqrt{1 - e^{2}} \right )
\nonumber \\
&=&   - ~e + \frac{e}{\sqrt{1-e^{2}}} \geq 0 ~.
\end{eqnarray}
Thus, $w(e)$ is an increasing function of eccentricity.
We obtain that $w(e)$ $\geq$ 0 for all $e$ $\in$ [0,1), since $w(e)$
is an increasing function of eccentricity and $w(0)$ $=$ 0.
Hence, the sign of the second term in the square braces in Eq. (23)
will depend on the sign of $I^{2}-S^{2}$.

For small values of the eccentricity we get that the secular time derivative
of eccentricity is, approximately, proportional only to the first
term multiplied by $I$ in the square braces in Eq. (23).

Eq. (24) yields that the argument of perihelion is constant for
the planar case $C$ $\equiv$ 0 and for the orbit orientation
characterized by $S$ $=$ 0. The longitude of the ascending node
and the inclination are constant in this case. Hence
differentiation of the second equation of Eqs. (27) with
respect to time gives
\begin{equation}\label{31}
\frac{dI}{dt} = - ~S ~\frac{d \omega}{dt} ~.
\end{equation}
Thus, $I$ is also constant in this case. Since $\sigma$
is a small number, the first term in square braces in Eq. (23)
is the dominant term on evolution eccentricity.

Now, let us consider the planar case ($C$ $\equiv$ 0) with $S$ $\neq$ 0.
The differentiation of the first of Eqs. (27) with respect
to time gives
\begin{equation}\label{32}
\frac{dS}{dt} = I ~\frac{d \omega}{dt} ~.
\end{equation}
This quantity can be averaged using Eq. (21). We get
\begin{equation}\label{33}
\left \langle \frac{dS}{dt} \right \rangle = I
\left \langle \frac{d \omega}{dt} \right \rangle ~.
\end{equation}
If $\sigma$ is a small number and $I$ is not close to zero,
then we can use the following approximations of Eqs. (23)-(24)
\begin{equation}\label{34}
\left \langle \frac{de}{dt} \right \rangle \approx
\frac{3 ~c_{D} ~\gamma_{H} ~v_{H}}{2} ~\sqrt{\frac{p}{\mu}} ~I ~,
\end{equation}
\begin{equation}\label{35}
\left \langle \frac{d \omega}{dt} \right \rangle \approx
- ~\frac{3 ~c_{D} ~\gamma_{H} ~v_{H}}{2} ~\sqrt{\frac{p}{\mu}} ~\frac{S}{e} ~.
\end{equation}
Inserting Eqs. (34)-(35) into Eq. (33) one can obtain
\begin{equation}\label{36}
\left \langle \frac{de}{dt} \right \rangle \approx - ~
\frac{e}{S} \left \langle \frac{dS}{dt} \right \rangle ~.
\end{equation}
This equation leads to the differential equation
\begin{equation}\label{37}
\frac{de}{e} \approx - ~\frac{dS}{S} ~,
\end{equation}
with the solution
\begin{equation}\label{38}
e \approx \frac{D}{\vert S \vert} ~.
\end{equation}
$D$ is a constant which can be determined from initial conditions.
Thus, eccentricity is close to its minimal value if the major axis
of the orbit is aligned with the direction of the hydrogen gas
velocity vector.

If we consider the system Eqs. (23)-(26) with $\sigma$ $=$ 0,
then we obtain $\langle dS/dt \rangle$ $=$ $-$ $3$ $c_{D}$
$\gamma_{H}$ $v_{H}$ $\sqrt{p / \mu}$ $I$ $S$ $/$ $(2e)$
from the first of Eqs. (27).
If we use Eq. (23) with $\sigma$ $=$ 0, then we get equation
$\langle de /dt \rangle$ $=$ $-$ $e$ $\langle dS /dt \rangle$ $/$ $S$.
This equation leads to the solution $e$ $=$ $D$ $/$ $\vert S \vert$ also
for this non-planar case. But approximation $\sigma$ $\approx$ 0 is
not allowed in Eqs. (24)-(26) since the first terms
in square braces multiplied by $C$ are multiplied by $e$, $\sin \omega$
and $\cos \omega$ which can be close to zero.

The case $\sigma$ $=$ 0 (the RHS of Eq. 1 is independent of the
particle's velocity -- constant force) is also significant for a motion
of a rocket in a central gravitational field with a constant-reaction
acceleration vector. We obtain the following important result:
if the constant acceleration can be considered as a perturbation
acceleration to the central gravitation and $S$ $\neq$ 0, then equation
$e$ $=$ $D$ $/$ $\vert S \vert$ holds during orbital motion
of the rocket. This result fills up the results
of Betelsky (1964) and Kunitsyn (1966).

\begin{figure*}[t]
\begin{center}
\includegraphics[width=0.85\textwidth]{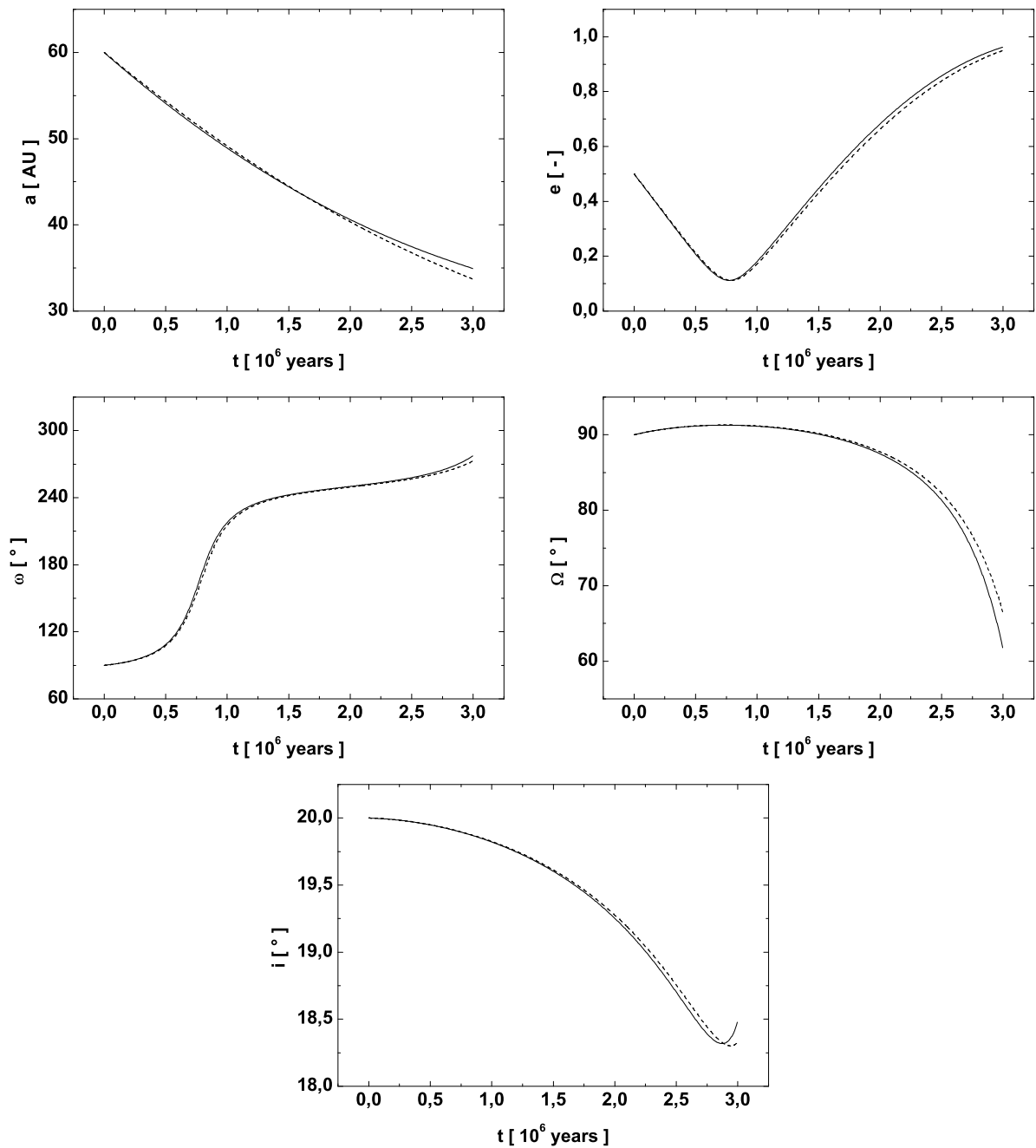}
\end{center}
\caption{Two evolutions of orbital elements of a dust particle with
$R$ $=$ 2 $\mu$m and mass density $\varrho$ $=$ 1 g/cm$^{3}$ under
the action of interstellar gas flow. Evolution depicted by a solid line
is calculated from the equation of motion. Evolution depicted
by a dashed line corresponds to Eqs. (22)-(26).}
\label{F4}
\end{figure*}

Parameters $S$, $I$ and $C$ determine position of the orbit with
respect to the hydrogen gas velocity vector. Therefore their time derivatives
are useful for description of evolution of the orbit position in space.
Putting Eqs. (24)-(26) into averaged time derivatives of
the quantities $S$, $I$ and $C$ defined by Eqs. (27), one obtains
\begin{eqnarray}\label{39-41}
\left \langle \frac{dS}{dt} \right \rangle &=& ~
      \frac{c_{D} ~\gamma_{H} ~v_{H} ~S}{2} ~\sqrt{\frac{p}{\mu}} ~
      \Biggl \{ - ~\frac{3I}{e} - \frac{\sigma}{v_{H}}
      \biggl [C^{2}
\nonumber \\
& &   - ~\frac{I^{2}}{e^{4}} \biggl ( e^{4} - 6e^{2} + 4 -
      4(1 - e^{2})^{3/2} \biggr ) \biggr ] \Biggr \} ~,\\
\left \langle \frac{dI}{dt} \right \rangle &=& ~
      \frac{c_{D} ~\gamma_{H} ~v_{H}}{2} ~\sqrt{\frac{p}{\mu}} ~
      \Biggl \{ - ~\frac{3eC^{2}}{1-e^{2}} + \frac{3S^{2}}{e} -
      \frac{\sigma I}{v_{H}} \biggl [C^{2}
\nonumber \\
& &   + ~\frac{S^{2}}{e^{4}} \biggl ( e^{4} - 6e^{2} + 4 -
      4(1 - e^{2})^{3/2} \biggr ) \biggr ] \Biggr \} ~,\\
\left \langle \frac{dC}{dt} \right \rangle &=& ~
      \frac{c_{D} ~\gamma_{H} ~v_{H} ~C}{2} ~\sqrt{\frac{p}{\mu}} ~
      \Biggl [ \frac{3eI}{1-e^{2}}
\nonumber \\
& &   + ~\frac{\sigma}{v_{H}}(S^{2} + I^{2}) \Biggr ] ~.
\end{eqnarray}
Eq. (39) yields for the planar case ($C$ $\equiv$ 0)
\begin{eqnarray}\label{42}
\left \langle \frac{dS}{dt} \right \rangle &=& ~
      \frac{c_{D} ~\gamma_{H} ~v_{H} ~S}{2} ~\sqrt{\frac{p}{\mu}} ~
      \Biggl [ - ~\frac{3I}{e}
\nonumber \\
& &   + ~\frac{\sigma I^{2}}{v_{H}e^{4}} \biggl ( e^{4} - 6e^{2} + 4 -
      4(1 - e^{2})^{3/2} \biggr ) \Biggr ]
\nonumber \\
&=&   \frac{c_{D} ~\gamma_{H} ~v_{H} ~S}{2} ~\sqrt{\frac{p}{\mu}} ~
      \biggr ( - ~\frac{3I}{e} + \frac{\sigma I^{2}}{v_{H}}
      k(e) \biggl ) ~.
\end{eqnarray}
Function $k(e)$ is a decreasing function of eccentricity for $e$ $\in$ (0,1).
This can be shown in a similar procedure as we used in Eq. (30) for
the function $w(e)$. The procedure must be used more than once for
the function $k(e)$. The function $k(e)$ obtains values from
$\lim_{e \to 0} k(e)$ $=$ $-$ 0.5 to $\lim_{e \to 1} k(e)$ $=$ $-$ 1,
for $e$ $\in$ (0,1). For the eccentricity $e$ $=$ 1 we obtain
$\lim_{e \to 1} \sigma$ $=$ $\infty$. Hence we use maximal value
of eccentricity $e_{m}$ for which Eq. (15) holds.
The first term in parenthesis in Eq. (42)
is minimal for $e$ $=$ $e_{m}$ for a given value of $I$.
Value of the parenthesis in Eq. (42) will be maximally affected by the
second term for $e$ $=$ $e_{m}$. The second term in parenthesis
yields for $e$ $=$ $e_{m}$
\begin{equation}\label{43}
\frac{\sigma I^{2}}{v_{H}} ~k(e_{m}) \geq - ~
\frac{\sigma I^{2}}{v_{H}} \geq - ~\sigma \vert I \vert \geq - ~
\vert I \vert \geq - ~3 ~\frac{\vert I \vert}{e_{m}} ~,
\end{equation}
since $\sigma$ $\ll$ 1 and $\vert I \vert$ $\leq$ $v_{H}$.
Thus, the sign of $\langle dS / dt \rangle$ depends only on
the sign of the first term in the parenthesis in Eq. (42).
Fig. 3 depicts unit vectors $\vec{e}_{RP}$ $=$ $\vec{e}_{R}(f=0)$
and $\vec{e}_{TP}$ $=$ $\vec{e}_{T}(f=0)$ in the orbital plane
of a particle with prograde orbit. The hydrogen gas velocity vector
$\vec{v}_{H}$ for the planar case lies in the orbital plane.
Direction of $\vec{v}_{H}$ is also shown.
The unit vector $\vec{e}_{RP}$ has direction and orientation
from the Sun to perihelion of the particle orbit. In Fig. 3a
$\vec{e}_{RP}$ lies in the first quadrant of the Cartesian coordinate
system with origin in the Sun and vertical axis aligned with
the direction of the hydrogen gas velocity vector. Both $S$ and $I$
are greater than 0, for these positions of the unit vectors
$\vec{e}_{RP}$ and $\vec{e}_{TP}$. Hence $\langle dS / dt \rangle$ is
negative. If $\vec{e}_{RP}$ lies in the second quadrant (Fig. 3b), then
$\langle dS / dt \rangle$ is positive. If
$\vec{e}_{RP}$ lies in the third quadrant (Fig. 3c),
then $\langle dS / dt \rangle$ is negative. Finally,
if $\vec{e}_{RP}$ lies in the fourth quadrant (Fig. 3d),
then $\langle dS / dt \rangle$ is positive, as it is
in the second quadrant. Therefore, the vector
$\vec{e}_{RP}$ rotates counterclockwise
in the first and the second quadrants, and, clockwise
in the third and the fourth quadrants.
If the vector $\vec{e}_{RP}$ is parallel with vertical axis in Fig. 3
($I$ $=$ 0 and $C$ $=$ 0), then Eq. (40) yields
$\langle dI/dt \rangle$ $>$ 0. Thus, positions of the vector
$\vec{e}_{RP}$ parallel with the vertical axis in Fig. 3 are not stable.
However, if $\vec{e}_{RP}$ is parallel with the horizontal axis
($S$ $=$ 0 and $C$ $=$ 0), then $\langle dS/dt \rangle$ $=$ 0 and
$\langle dI/dt \rangle$ $=$ 0. Thus, $\vec{e}_{RP}$ parallel with the
horizontal axis yields stable positions of the vector $\vec{e}_{RP}$.
The stable position of the vector $\vec{e}_{RP}$
parallel with the horizontal axis and directed to the left in Fig. 3 is of
theoretical importance, only. In reality, no particles should
be observed with perihelia in this direction. However, all unit vectors
$\vec{e}_{RP}$ of particles in prograde orbits will approach the
right direction in Fig. 3. For retrograde orbits we have to use
transformation $S$ $\rightarrow$ $S$, $I$ $\rightarrow$ $-$ $I$. We obtain
$\langle dS/dt \rangle$ $\rightarrow$ $-$ $\langle dS/dt \rangle$.
Hence, all unit vectors $\vec{e}_{RP}$ of particles in retrograde orbits
will approach the left direction in Fig. 3. This result was obtained,
in a different way, also by Scherer (2000). Scherer states that the
approaches of unit vectors $\vec{e}_{RP}$ to one direction holds also for
the non-planar case. However, the statement of Scherer is incorrect,
in general. If $I$ $=$ 0 and $C$ $\neq$ 0, then Eq. (39) implies that
$\langle dS / dt \rangle$ is proportional to $SC^{2}$ and not to $SI$.
This leads to a behavior which differs from the behavior discussed above
using Fig. 3.

Scherer (2000, p. 332) furthermore states that orbital plane
under the effect of the interstellar gas flow will
be rotated into a plane coplanar to the flow vector $\vec{v}_{H}$,
independent of initial position of the orbital plane. We showed,
using numerical integrations, that the statement of Scherer is not correct.
Retrograde orbits were not discussed by Scherer.
We found an interesting orbit behavior. It depends on
the orbit orientation with respect to the
hydrogen gas velocity vector $\vec{v}_{H}$. If $\vec{v}_{H}$
lies in a plane $i$ $=$ 0 and $\vec{v}_{H}$
is perpendicular to the vector $\vec{e}_{RP}$
(in this case $S$ $=$ 0 and $\langle dS/dt \rangle$ $=$ 0), then the
interstellar gas flow can change prograde orbit into a retrograde one
(even more times for one particle) and inclination does not approach
the value $i$ $=$ 0.

\section{Numerical results}

\subsection{Comparison of the numerical solution of equation
of motion and the solution of Eqs. (22)-(26)}

We numerically solved Eq. (3) and the system of differential equations
represented by Eqs. (22)-(26). Solutions are compared in Fig. 4.
We assumed that the direction of the interstellar gas velocity vector
is identical to the direction of the velocity of the interstellar dust
particles entering the Solar System. The interstellar dust particles
enter the Solar System with a speed of about $v_{\infty}$ $=$ 26 km/s
(Landgraf et al. 1999) and they are arriving from direction of
$\lambda_{ecl}$ $=$ 259$^{\circ}$ (heliocentric ecliptic longitude)
and $\beta_{ecl}$ $=$ 8$^{\circ}$ (heliocentric ecliptic latitude)
(Landgraf 2000). Thus, components of the velocity
in the ecliptic coordinates with x-axis aligned toward
the equinox are $\vec{v}_{H}$ $=$ $-$ 26 km/s
($\cos(259^{\circ}) ~\cos(8^{\circ})$, $\sin(259^{\circ}) ~\cos(8^{\circ})$,
$\sin(8^{\circ})$). As an initial conditions for a dust particle with $R$ $=$
2 $\mu$m and mass density $\varrho$ $=$ 1 g/cm$^{3}$
we used $a_{in}$ $=$ 60 AU, $e_{in}$ $=$ 0.5,
$\omega_{in}$ $=$ 90$^{\circ}$, $\Omega_{in}$ $=$
90$^{\circ}$, $i_{in}$ $=$ 20$^{\circ}$ for Eqs. (22)-(26).
The initial true anomaly of the dust particle was
$f_{in}$ $=$ 180$^{\circ}$ for Eq. (3).
Fig 4. shows that the obtained evolutions are in a good agreement.
Evolutions begin separate as the eccentricity
approaches 1. This is caused by the fact that approximation
$\sigma$ $\ll$ 1, see Eq. (15), does not hold for large
eccentricities. Detailed numerical solution of the equation of motion
(Eq. 3) yields that the secular semi-major axis is a decreasing function
of time also when eccentricity approaches 1.

We can summarize, on the basis of the previous paragraph.
If there is no other force, besides solar gravity and the flux
of interstellar gas, then the semi-major axis of an interplanetary
dust particle decreases and the particle can hit the Sun. However,
the particle can hit the Sun also by another possibility: particle's
eccentricity increases to 1. These mathematical possibilities
probably do not occur in reality, since other forces can
act on the dust particle and the interstellar gas is ionized below
the heliocentric distance of about 4 AU.

Let us return, once again, to the planar case ($C$ $\equiv$ 0) in which
$S$ $=$ 0 and the dominant term in the square braces
in Eq. (23), the term (3/2) $I$, is negative. Numerical integration
of Eq. (3) shows that if the eccentricity decreases to 0, then
the argument of perihelion $\omega$ "shifts" its value
to the value $\omega$ $+$ $\pi$. This means that the negative
value of $I$ changes to positive and the eccentricity begins to increase
with the same slope.

The approximative solution represented by Eq. (38)
is in a good agreement with the detailed numerical solution
of Eqs. (22)-(26) for the planar case with $S$ $\neq$ 0.
This holds for the whole time interval, also for $I$ close to zero.
Eq. (38) holds, approximately, also for
the non-planar evolution depicted in Fig. 4. In this case $i$ is close
to zero, $v_{H}$ $\approx$ $v_{HY}$ and $\Omega$ $\approx$ 90$^{\circ}$
at the eccentricity minimum. Eq. (38) gives $e$ $\approx$ $D$ $/$
$\vert v_{H} \cos \omega \vert$. The evolutionary minimum of eccentricity
occurs for the case when $\omega$ is close to 180$^{\circ}$.
This is in accordance with Eq. (38).

\begin{figure}[t]
\begin{center}
\includegraphics[width=0.45\textwidth]{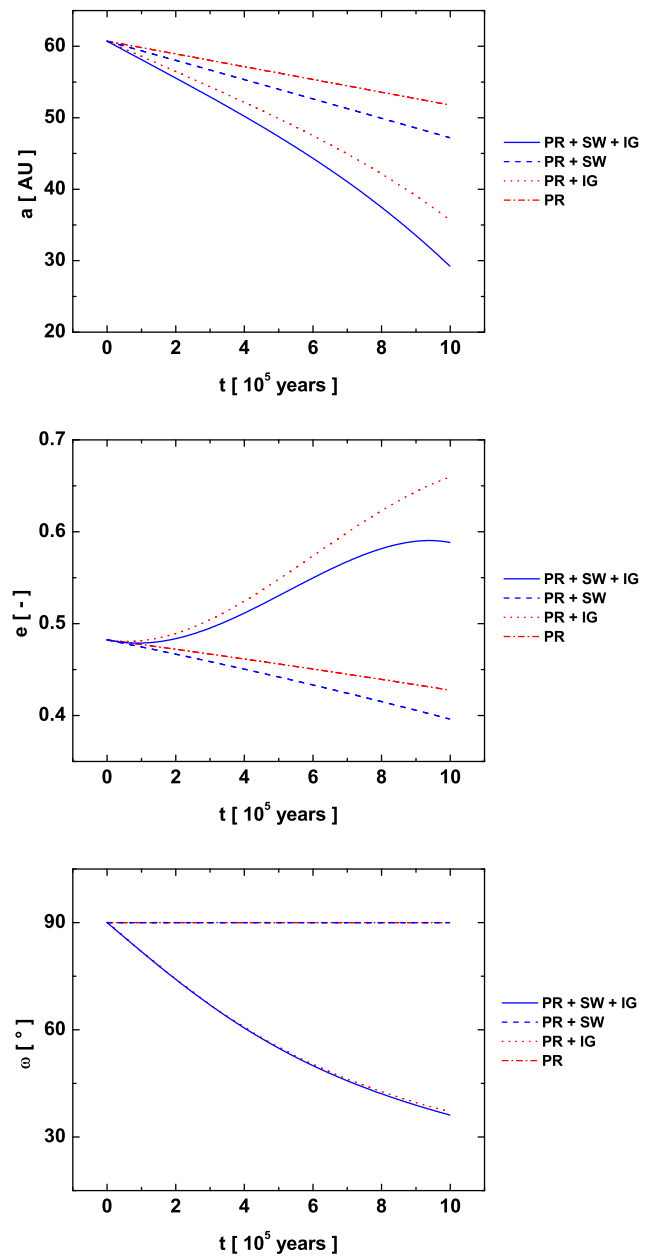}
\end{center}
\caption{Orbital evolution of dust particle with $R$ $=$ 2 $\mu$m,
mass density $\varrho$ $=$ 1 g/cm$^{3}$ and $\bar{Q}'_{pr}$ $=$ 0.75 under
the action of the Poynting-Robertson effect (PR), the radial solar wind (SW)
and the flow of interstellar gas (IG).}
\label{F5}
\end{figure}

\subsection{Comparing of influences of interstellar gas flow,
Poynting-Robertson effect and radial solar wind on dynamics of dust particles}

We have considered only the effect of interstellar gas flow, up to now.
In reality, some other non-gravitational effects play non-negligible role.
Thus, we want to compare the effect of the interstellar gas flow with
the other effects influencing dynamics of dust grains in the Solar System.
For this purpose we included the Poynting-Robertson effect (P-R effect)
and the radial solar wind into the equation of motion.
The P-R effect is electromagnetic radiation
pressure force acting on a spherical particle (Kla\v{c}ka 2004;
arXiv:astro-ph/0807.2915; arXiv:astro-ph/0904.0368).
Equation of motion for the P-R effect,
the effect of the radial solar wind and the effect of gas flow
has the form (e.g. Kla\v{c}ka et al. 2009b)
\begin{eqnarray}\label{44}
\frac{d \vec{v}}{dt} &=& - ~\frac{\mu \left ( 1 - \beta \right )}{r^{2}} ~
      \vec{e}_{R}
\nonumber \\
& &   - ~\beta ~\frac{\mu}{r^{2}}
      \left ( 1 + \frac{\eta}{\bar{Q}'_{pr}} \right )
      \left (\frac{\vec{v} \cdot \vec{e}_{R}}{c} ~\vec{e}_{R}
      + \frac{\vec{v}}{c} \right )
\nonumber \\
& &   - ~c_{D} ~\gamma_{H} ~
\vert \vec{v} - \vec{v}_{H} \vert ~\left ( \vec{v} - \vec{v}_{H} \right ) ~,
\end{eqnarray}
where decrease of particle's mass (corpuscular sputtering)
and higher orders in $\vec{v} / c$ are neglected.
$c$ is the speed of light in vacuum. Parameter $\beta$
is defined as the ratio of the radial component of the electromagnectic
radiation pressure force and the gravitational force between
the Sun and the particle in rest with respect to the Sun:
\begin{equation}\label{45}
\beta = \frac{3 L_{\odot} \bar{Q}'_{pr}}{16 \pi c G M_{\odot} R \varrho}
      \doteq 5.763 \times 10^{-4} ~
      \frac{\bar{Q}'_{pr}}{R ~[\mbox{m}] ~\varrho ~[\mbox{kg/m}^{3}]} ~.
\end{equation}
$L_{\odot}$ is the solar luminosity, $L_{\odot}$ $=$ 3.842 $\times$
10$^{26}$ W (Bahcall 2002), $\bar{Q}'_{pr}$ is the dimensionless efficiency
factor for radiation pressure integrated over the solar spectrum
and calculated for the radial direction ($\bar{Q}'_{pr}=1$ for
a perfectly absorbing sphere) and $\varrho$ is mass density of the particle.
$\eta$ is the ratio of solar wind energy to electromagnetic solar energy,
both radiated per unit of time
\begin{equation}\label{46}
\eta = \frac{4 \pi r^{2} u}{L_{\odot}} ~\sum_{i = 1}^{N} n_{i} m_{i} c^{2}~,
\end{equation}
where $u$ is the speed of the solar wind, $u$ $=$ 450 km/s,
$m_{i}$ and $n_{i}$, $i$ $=$ 1 to $N$, are masses and concentrations
of the solar wind particles at a distance $r$ from the Sun,
$\eta$ $=$ 0.38 for the Sun (Kla\v{c}ka et al. 2009b).
Four numerical integrations of Eq. (44) are shown in Fig. 5.
We used dust particle with $R$ $=$ 2 $\mu$m,
mass density $\varrho$ $=$ 1 g/cm$^{3}$ and
$\bar{Q}'_{pr}$ $=$ 0.75. We took into account only
planar case when the interstellar gas velocity lies in
the orbital plane of the dust particle ($C$ $=$ 0),
for the sake of simplicity. Initial position is
$\vec{r}_{in}$ $=$ (0, $-$90 AU, 0) and initial velocity vector
is $\vec{v}_{in}$ $=$ (2 km/s, 0, 0). Orbital evolution is given by
evolution of orbital elements calculated for the case when a central
acceleration is defined by the first Keplerian term in Eq. (44), namely
$-\mu (1-\beta)\vec{e}_{R}/r^{2}$. The evolution depicted by the dash-dotted
line is for the P-R effect alone ($\gamma_{H}$ $=$ 0,
$\eta$ $=$ 0 in Eq. 44). The evolution depicted by the dotted line is for
the P-R effect with the flow of interstellar gas
($\eta$ $=$ 0 in Eq. 44). The evolution depicted by the dashed line is for
the P-R effect and the radial solar wind ($\gamma_{H}$ $=$ 0
in Eq. 44). Finally, the evolution depicted by the solid line holds for
the case when all three effects act together.

The evolution of semi-major axis depicted in Fig. 5 shows that
the flow of interstellar gas is more important than the radial
solar wind, as for the effects on the dynamics of the dust
particle.

The secular eccentricity is always a decreasing
function of time for the P-R effect and the radial solar wind
(e.g., Wyatt \& Whipple 1950, Kla\v{c}ka et al. 2009b).
The growth in eccentricity depicted in Fig. 5 is due to the
interstellar gas. Fast decrease of the semi-major axis in Fig. 5
may also be, at least partially, caused by the fact that higher
eccentricities decrease the value of $\langle da/dt \rangle _{PR}$
and the P-R effect becomes stronger. The secular evolution of
eccentricity can be also an increasing function of time if the flow
of interstellar gas is taken into account. We have
\begin{eqnarray}\label{47}
\left \langle \frac{de_{\beta}}{dt} \right \rangle &=&
      - ~\frac{5}{2} ~\beta ~\left (1 + \frac{\eta}{\bar{Q}'_{pr}} \right ) ~
      \frac{\mu}{c} ~\frac{e_{\beta}}{a^{2}_{\beta}(1 - e^{2}_{\beta})^{1/2}}
\nonumber \\
& &   + ~c_{D} ~\gamma_{H} ~
      v_{H} ~\sqrt{\frac{p_{\beta}}{\mu (1 - \beta)}} ~
\nonumber \\
& &   \times ~\Biggl [\frac{3I_{\beta}}{2} +
      \frac{\sigma (I^{2}_{\beta} - S^{2}_{\beta})(1 - e_{\beta}^{2})}
      {v_{H}e^{3}_{\beta}}
\nonumber \\
& &   \times ~\left (1 - \frac{e^{2}_{\beta}}{2} -
      \sqrt{1 - e^{2}_{\beta}} \right ) \Biggr ] ~,
\end{eqnarray}
if also Eq. (23) is used. The subscript $\beta$ denotes
that the central acceleration $-\mu (1-\beta)\vec{e}_{R}/r^{2}$
is used for calculation of the osculating orbital elements.
We remind that the transformation $\mu$ $\rightarrow$
$\mu (1 - \beta)$ has to be done on the RHS sides
of Eqs. (20)-(26). If we use definition of the osculating
orbital elements, then the physical central acceleration is
given by the gravitational acceleration of the Sun,
$-\mu \vec{e}_{R}/r^{2}$. In this case, the secular evolution of
eccentricity is given by Eq. (103) in Kla\v{c}ka (2004),
assuming that $e_{\beta}$ is calculated from Eq. (47).

If optical properties of the dust particle are
constant, then the secular time derivative of the argument of perihelion
equals to zero for the P-R effect and the radial solar wind
(Kla\v{c}ka et al. 2007, 2009b). If the flow of interstellar gas
is included into the equation of motion, then, even in the planar case,
the secular time derivative of the argument of perihelion
may not be equal to zero, in general (see Fig. 5).

The evolutions of eccentricity and argument of perihelion shown in Fig. 5 are
significantly affected by the flow of interstellar gas.

\begin{figure*}[t]
\begin{center}
\includegraphics[width=0.45\textwidth]{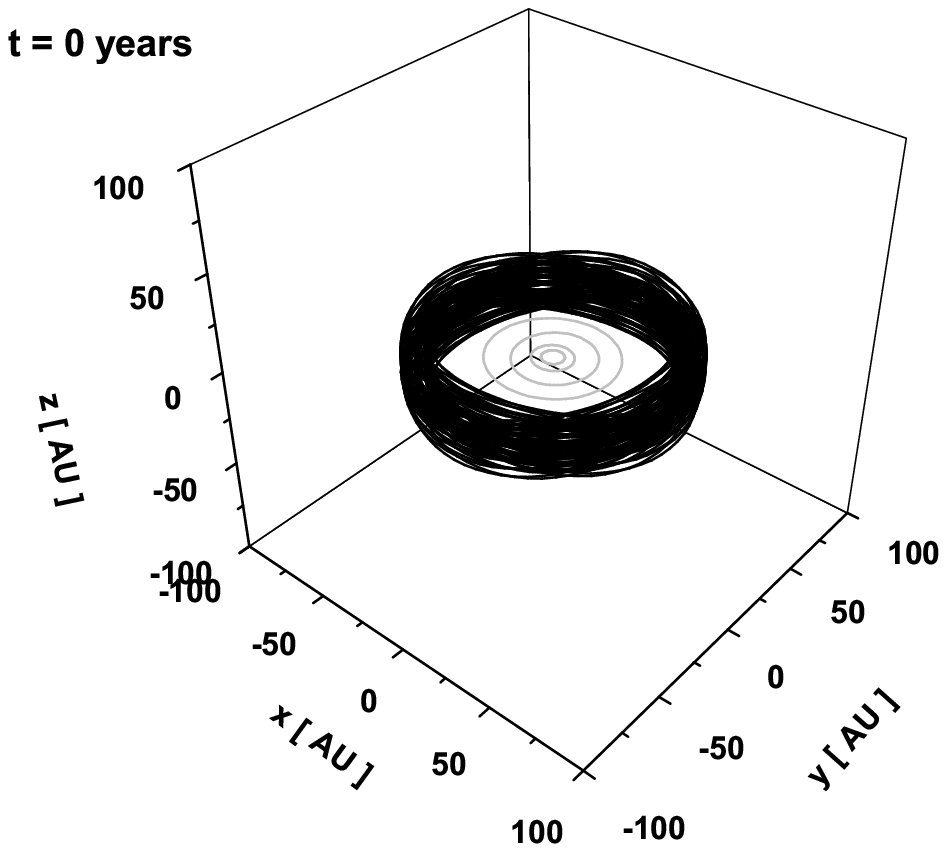}
\includegraphics[width=0.45\textwidth]{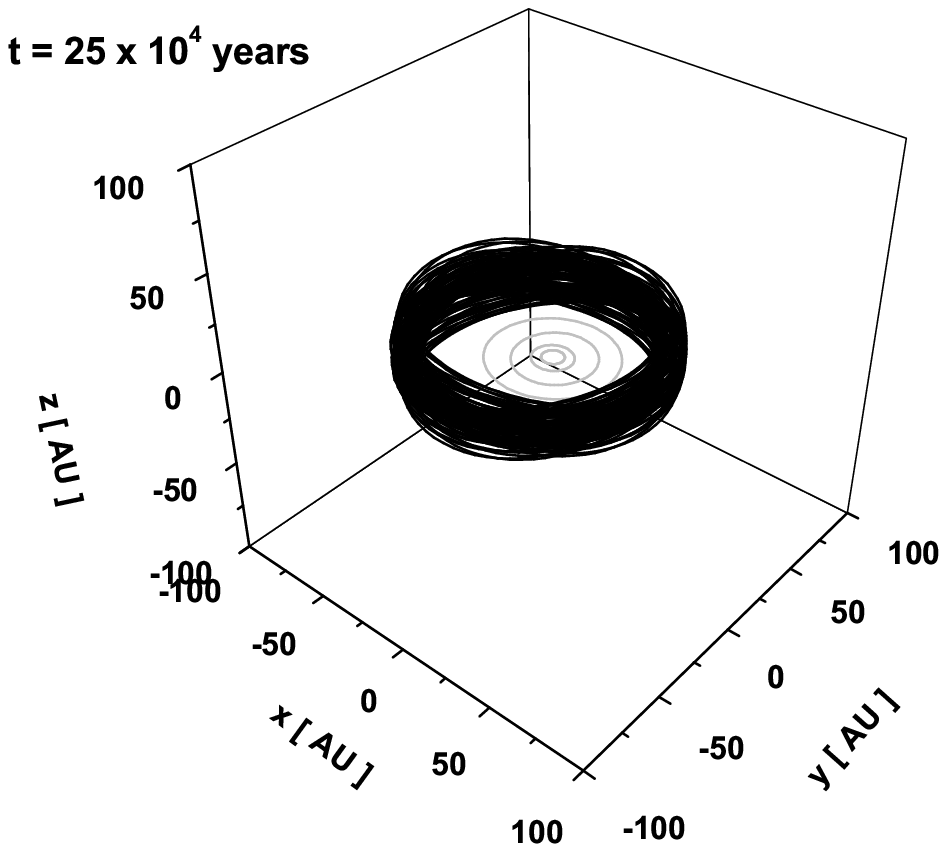}
\includegraphics[width=0.45\textwidth]{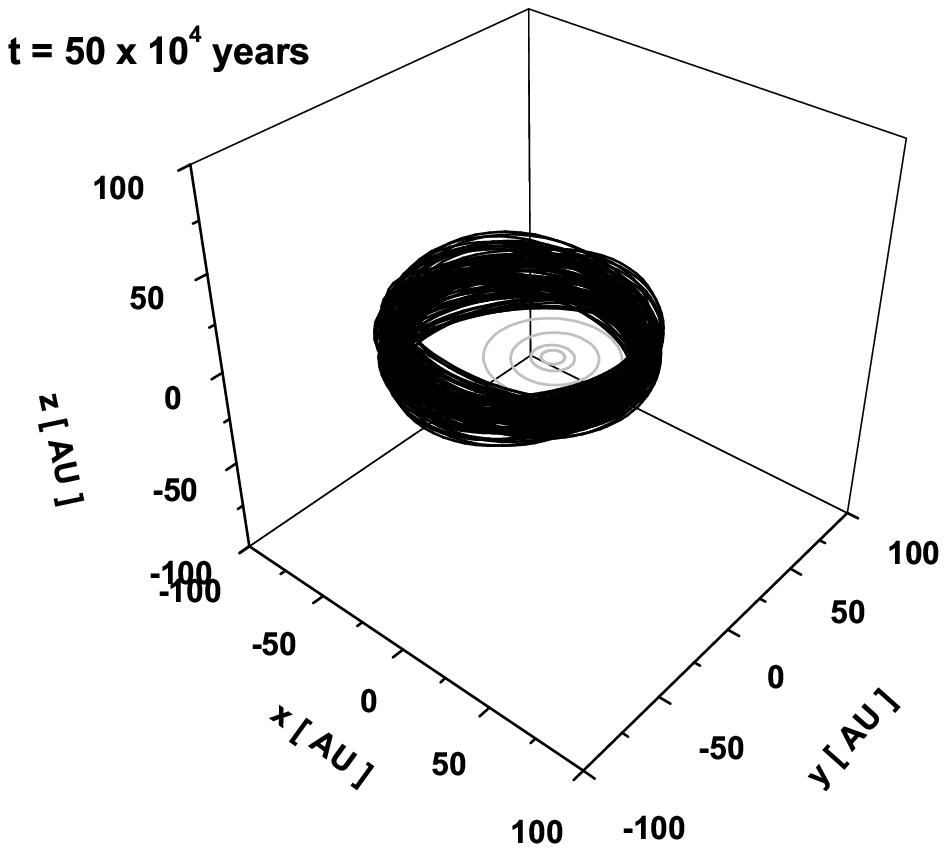}
\includegraphics[width=0.45\textwidth]{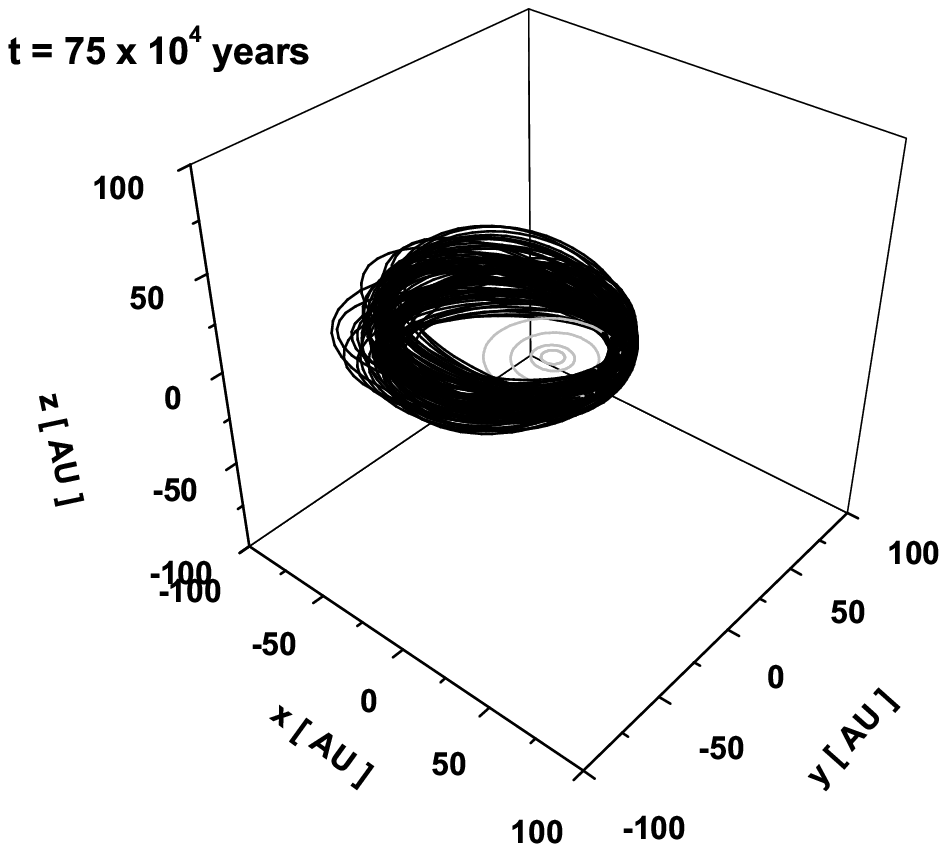}
\end{center}
\caption{Time evolution of ring of dust particles with $R$ $=$ 2 $\mu$m,
$\varrho$ $=$ 1 g/cm$^{3}$ and $\bar{Q}'_{pr}$ $=$ 0.75 in the zone
of the Edgeworth-Kuiper belt. The ring becomes eccentric in less than
10$^{6}$ years due to the interstellar neutral gas. Orbits of the particles
are shown with black color and orbits of the planets are shown in gray.}
\label{F6}
\end{figure*}

\begin{figure*}[t]
\begin{center}
\includegraphics[width=0.85\textwidth]{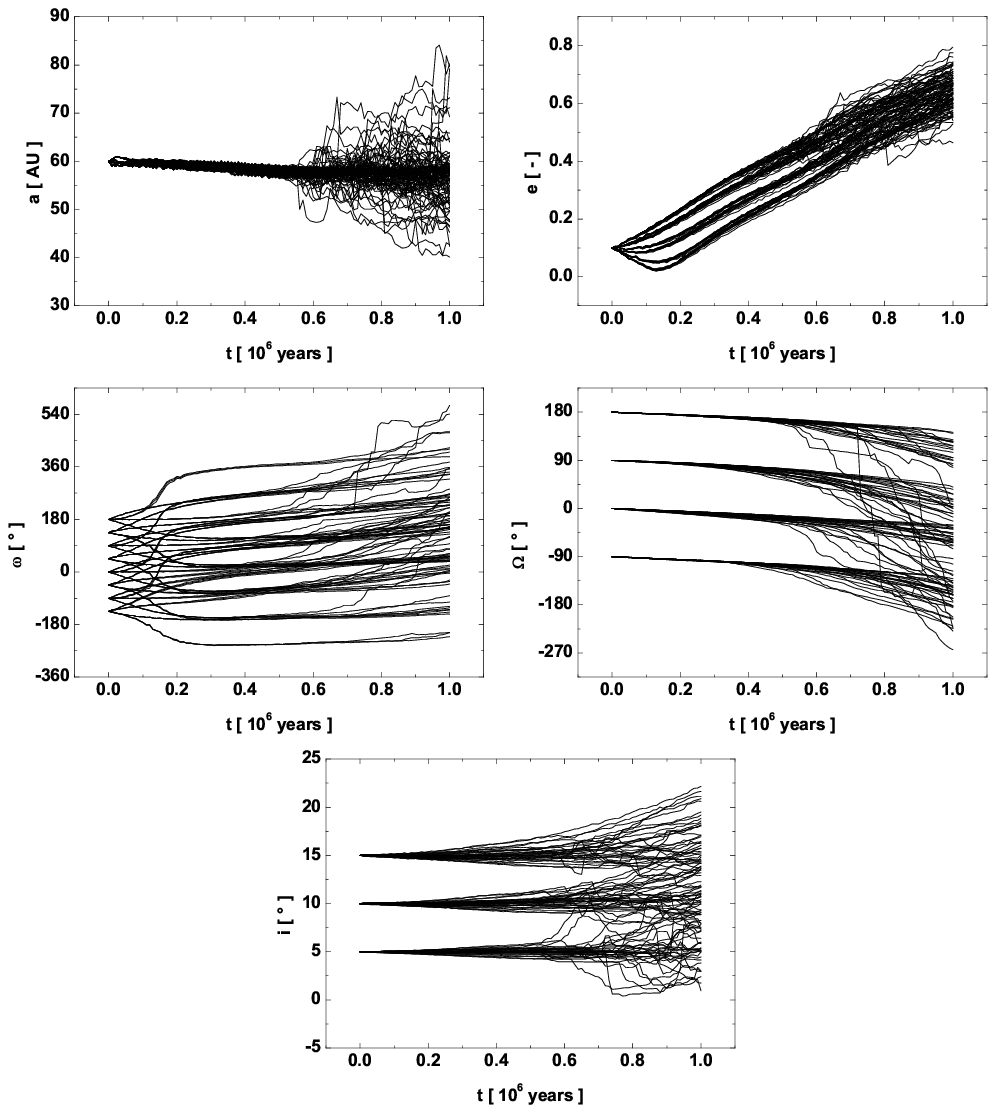}
\end{center}
\caption{Evolution of orbital elements of dust particles in
the Edgeworth-Kuiper belt zone during 96 numerical solutions depicted
in Fig. 6. Initial values of orbital elements are:
$a_{in}$ $=$ 60 AU, $e_{in}$ $=$ 0.1, $\omega_{in}$ $\in$ \{0$^{\circ}$,
45$^{\circ}$, 90$^{\circ}$, ..., 270$^{\circ}$, 315$^{\circ}$\},
$\Omega_{in}$ $\in$ \{0$^{\circ}$, 90$^{\circ}$, 180$^{\circ}$,
270$^{\circ}$\}, $i_{in}$ $\in$ \{5$^{\circ}$, 10$^{\circ}$, 15$^{\circ}$\}
and $f_{in}$ $=$ 0$^{\circ}$. Evolution during the first 750 000 years is
influenced mainly by interstellar gas, and, later on, mainly
by gravitation of planets (see text).}
\label{F7}
\end{figure*}

\subsection{Dust ring in the Edgeworth-Kuiper belt zone}

Real situation in the Edgeworth-Kuiper belt zone may be much more
complicated than the situation discussed in Secs. 4.1 and 4.2.
In particular, gravitation of planets may have an important
influence on dynamics of dust in the zone. For this reason we included
gravitation of four major planets into the final equation of motion.
Observations from Helios 2 during its first solar mission in 1976
(Bruno et. al. 2003) show that the angle between the radial
direction and the direction of the solar wind velocity does
not significantly depend on heliocentric distance.
If the value of this angle is approximately constant,
then the non-radial solar wind can also have an important
influence on dynamics of dust in outer parts of the Solar System.
We took into account the non-radial solar
wind with constant value of the angle. Influence of precession of
the rotational axis of the Sun on the non-radial solar wind was also
considered. Hence, equation of motion of the dust particle has the form
\begin{eqnarray}\label{48}
\frac{d \vec{v}}{dt} &=& - ~\frac{\mu}{r^{2}} ~\vec{e}_{R}
\nonumber \\
& &   + ~\beta ~\frac{\mu}{r^{2}}
      \left [ \left ( 1 - \frac{\vec{v} \cdot \vec{e}_{R}}{c} \right ) ~
      \vec{e}_{R} - \frac{\vec{v}}{c} \right ]
\nonumber \\
& &   - ~\frac{\beta ~\eta }{\bar{Q}'_{pr} ~c ~u} ~\frac{\mu}{r^{2}}
\vert \vec{v} - \vec{u} \vert ~\left ( \vec{v} - \vec{u} \right )
\nonumber \\
& &   - ~c_{D} ~\gamma_{H} ~
\vert \vec{v} - \vec{v}_{H} \vert ~\left ( \vec{v} - \vec{v}_{H} \right )
\nonumber \\
& &   - ~\sum_{i=1}^{4} ~\frac{GM_{i}}
      {\vert \vec{r} - \vec{r_{i}} \vert^{3}} ~
      (\vec{r} - \vec{r_{i}})
\nonumber \\
& &   - ~\sum_{i=1}^{4} ~\frac{GM_{i}}{\vert \vec{r_{i}} \vert^{3}} ~
      \vec{r_{i}} ~,
\end{eqnarray}
where $\vec{u}$ is solar wind velocity vector, $M_{i}$ are masses of
the planets and $\vec{r}_{i}$ are position vectors of the planets with
respect to the Sun. The non-radial solar wind velocity vector was calculated
from equation
\begin{equation}\label{49}
\vec{u} = u \left (\vec{e}_{R} \cos \varepsilon +
\frac{\vec{k} \times \vec{e}_{R}}
{\vert \vec{k} \times \vec{e}_{R} \vert} \sin \varepsilon \right ) ~,
\end{equation}
where $\varepsilon$ is the angle between the radial direction and
the direction of the solar wind velocity and $\vec{k}$ is unit vector with
direction/orientation corresponding to the direction/orientation
of solar rotation angular velocity vector.
Vector $\vec{k}$ for a given time can be calculated from
\begin{eqnarray}\label{50}
\vec{k} &=& (\sin \Omega_{s} \sin i_{s},
      - \cos \Omega_{s} \sin i_{s}, \cos i_{s}) ~,
\nonumber \\
      i_{s} &=& 7^{\circ} 15', ~~ \Omega_{s} = 73^{\circ} 40' + 50.25'' ~
      (t [ \mbox{years} ] - 1850) ~.
\end{eqnarray}
While Eqs. (49)-(50) are consistent with Kla\v{c}ka (1994) and
Abalakin (1986), the value of $\varepsilon$,
$\varepsilon$ $=$ 2.9$^{\circ}$, used in our numerical calculations,
is in accordance with Bruno et al. (2003) and Kla\v{c}ka et al. (2007).
The observed neutral hydrogen gas velocity vector
in the ecliptic coordinates with $x$-axis aligned toward the equinox is
$\vec{v}_{H}$ $=$ $-$ 26 km/s ($\cos(259^{\circ}) ~\cos(8^{\circ})$,
$\sin(259^{\circ}) ~\cos(8^{\circ})$, $\sin(8^{\circ})$).
As for the initial conditions of dust particles we did not use
random positions and velocities. We assumed that putative dust
ring in the Edgeworth-Kuiper belt has approximate circular shape and contains
lot of particles with approximately equal optical properties.
We assumed that the ring contains such large amount of particles
that one can choose, approximately, a given value of semi-major axis and
a given radius of the particles. Accelerations caused by the P-R effect,
the solar wind and the interstellar neutral hydrogen gas are
inversely proportional to particle's radius and mass density. Therefore,
a large particle is less influenced by the non-gravitational effects
than a small particle of the same mass density. The evolution
of the large particle under action of the non-gravitational effects
is slower than the evolution of the small dust grain. We used
uniformly distributed initial values of the argument of perihelion
and longitude of the ascending node. Furthermore we assumed that particles
in the ring orbit prograde in low inclination orbits. We used particles with
$R$ $=$ 2 $\mu$m, $\varrho$ $=$ 1 g/cm$^{3}$ and $\bar{Q}'_{pr}$ $=$ 0.75.
Exact initial values of orbital elements were
$a_{in}$ $=$ 60 AU, $e_{in}$ $=$ 0.1, $\omega_{in}$ $\in$ \{0$^{\circ}$,
45$^{\circ}$, 90$^{\circ}$, ..., 270$^{\circ}$, 315$^{\circ}$\},
$\Omega_{in}$ $\in$ \{0$^{\circ}$, 90$^{\circ}$, 180$^{\circ}$,
270$^{\circ}$\}, $i_{in}$ $\in$ \{5$^{\circ}$, 10$^{\circ}$, 15$^{\circ}$\}
and $f_{in}$ $=$ 0$^{\circ}$. Hence, we obtained
8 $\times$ 4 $\times$ 3 $=$ 96 individual orbits.
Results of numerical solutions of Eq. (48) are depicted in Figs. 6 and 7.
Fig. 6 depicts evolution of the dust ring viewed from
perspective. Orbits of the planets are also shown.
Time span between various pictures in Fig. 6 is 250 000 years.
As we can see, the ring becomes more and more eccentric because of
a fast increase of eccentricity caused by the interstellar gas flow
(see also eccentricity evolution in Fig. 7). Perihelia
of orbits are shifted in accordance with the behavior
discussed in Fig. 3. This is caused by the facts that
influence of interstellar gas is dominant and the solved problem
is almost coplanar. The term multiplied by $C^{2}$ in Eq. (39) does not have
large influence on the first term in the curly
braces in Eq. (39), in almost coplanar case. Therefore, the
lines connecting the Sun with the perihelia of particles orbits are
approaching the direction perpendicular to the interstellar gas velocity
vector. Time evolution of the orbital elements of the dust particles
in the ring, considered in Fig. 6, is depicted in Fig. 7.
Due to the approach of the perihelia to one direction, the evolutions
of the argument of perihelion $\omega$, beginning with a given initial value
$\omega_{in}$, are divided into four branches. Each of them corresponds
to one initial value of the ascending node. If the time is less than
750 000 years, then: i) the concentration of the particles
in the ring is smallest in the direction (from the Sun) into which
perihelia of the orbits are approaching, and, ii) the
concentration of the particles is greatest in exactly opposite direction.
If the time is greater than 750 000 years, then the orbits of particles
in the dust ring are getting close to the orbits of the planets
due to the increase of particles eccentricities.
The situation after 750 000 years can be seen in Fig. 7.
The dust particles with $R$ $=$ 2 $\mu$m, $\varrho$ $=$ 1 g/cm$^{3}$
and $\bar{Q}'_{pr}$ $=$ 0.75 are characterized by the value $\beta$ $\approx$
0.216 (see Eq. 45). For this value of $\beta$, one obtains
$a$ $\approx$ 57.7 AU for the location of the exterior mean motion
3/1 resonance with Neptune; it follows from $a$ $=$ $a_{P}$
$(1 - \beta)^{1/3}$ $(3/1)^{2/3}$, where $a_{P}$ is semi-major
axis of Neptune. We can see, from the evolution of semi-major axis
in Fig. 7, that the secular semi-major axis is a decreasing function
of time during the first 750 000 years. Thus the semi-major axis
can evolve from an initial value of 60 AU to the location close to
the mean-motion 3/1 resonance. Particles are influenced
both by the vicinity of Neptune orbit and the exterior mean motion
3/1 resonance with Neptune. Evolution during the first 750 000 years
is influenced mainly by the neutral interstellar hydrogen gas and,
later on, mainly by the gravitation of planets.

Inclusion of the P-R effect, the non-radial solar wind and the interstellar
gas into the equation of motion of the dust particle without planets can
stabilize the particle's orbit. The stabilization is characterized by stable
values of orbital elements. This stabilization is discussed in
Kla\v{c}ka et al. (2009a) for $\sin \varepsilon$ $=$ 0.05. The process
of stabilization requires about 1 $\times$ 10$^{8}$ years for the dust
particle with $\beta$ $=$ 0.01. This time is not very sensitive to
the efficiency factor for radiation pressure $\bar{Q}'_{pr}$. However, for
lower values of $\bar{Q}'_{pr}$ the stabilization occur with larger
probability, because stabilization effect of the non-radial
solar wind is stronger (see Eq. 48). If also planets
are considered in the equation of motion, then the stabilizing value
of dust particle eccentricity is usually sufficiently high to get
the particle close to one of the planets during a long time span.
As a consequence, gravitation of the planet can change orbit of
the particle and cancel the stabilization process.

\section{Conclusion}

We investigated orbital evolution of a dust grain under the action of
interstellar gas flow. We presented secular time derivatives of the grain's
orbital elements for arbitrary orientation of the orbit with respect to
the velocity vector of the interstellar gas, which is a generalization of
several results presented in Kla\v{c}ka et al. (2009a). The secular time
derivatives were derived using assumptions that the acceleration caused by
the interstellar gas flow is small in comparison with gravitation
of a central object (the Sun), eccentricity of the orbit is not close to 1
and the speed of the dust particle is small in comparison with the speed of
the interstellar gas. These assumptions lead to secular decrease
of semi-major axis $a$ of the particle. The secular time derivative
of the semi-major axis is negative and proportional to $a$.
This result is not in accordance with Scherer (2000) who has stated
that the semi-major of the particle increases exponentially. Scherer's
statement is incorrect and our analytical result is confirmed by our
detailed numerical integration of equation of motion (see also Fig. 4).

If the hydrogen gas velocity vector $\vec{v}_{H}$ lies
in the particle's orbital plane and the major axis of the orbit
is not perpendicular to $\vec{v}_{H}$,
then the product of (secular) eccentricity and magnitude
of the radial component of $\vec{v}_{H}$ measured
in perihelion is, approximately, constant during orbital evolution.

We considered simultaneous action of the P-R effect,
the radial solar wind and the interstellar gas flow.
Numerical integrations showed that the action
of the flow of interstellar gas can be more important
than the action of the electromagnetic and corpuscular radiation
of the Sun, as for the motion of dust particles orbiting the Sun
in outer parts of the Solar System (see Fig. 5).
Physical decrease of semi-major axis can be more than 2-times greater
than the value produced by the P-R effect and radial solar wind.
The evolution of eccentricity can also be an increasing function
of time when we consider the P-R effect and the radial solar wind
together with the flow of the neutral interstellar gas. This
is also relevant difference from the action of the P-R effect
and the radial solar wind when secular decrease of eccentricity
occurs. Simultaneous action of all three effects yields that the secular
time derivative of the argument of perihelion may not be equal
to zero, in general.

Gravitation of four major planets was also directly added
into the equation of motion, see Eq. (48). This access correctly
describes capture of dust grains into mean motion resonances
with the planets. Our physical approach differs from the
Scherer's approach (Scherer 2000), who has used some kind of
secular access to gravitational influence of the planets.

Assumption on an existence of dust ring in the zone of
the Edgeworth-Kuiper belt is in contradiction with rapid increase
of eccentricity of the ring due to an acceleration caused by the
interstellar gas flow. Speed of the eccentricity increase (time
derivative of eccentricity) is roughly inversely proportional
to the particle's size and mass density.
As the eccentricity of the particles increases, the particles
approach the planets. The particles in the ring are under the
gravitational influence of the planets. The particles evolve also in
semi-major axis and they can be temporarily captured into a mean motion
resonance. The particles can remain in chaotic orbits between orbits of
the planets, or, the particles are ejected to high eccentric
orbits due to close encounters with one of the planets.
Only particles with greater size and mass density should
survive in the dust ring for a long time.

A relevant result of the paper is that equation of motion in the
form of Eq. (44) (or, Eq. 48) and Eqs. (45)-(46) have to be used in
modeling of orbital evolution of dust grains in the Solar System. The
influence of the fast interstellar neutral gas flow might not be ignored
in general investigations on evolution of dust particles in
the zone of the Edgeworth-Kuiper belt.

\acknowledgements{The paper was supported by the Scientific Grant Agency VEGA
grant No. 2/0016/09.}

\end{document}